\documentclass[conference]{IEEEtran}
\IEEEoverridecommandlockouts

\usepackage{cite}
\usepackage{amsmath,amssymb,amsfonts}
\usepackage{algorithm}
\usepackage{algorithmic}
\usepackage{graphicx}
\usepackage{multirow}
\usepackage{textcomp}
\usepackage{xcolor}
\usepackage{booktabs}
\usepackage{url}
\usepackage{eso-pic}
\def\BibTeX{{\rm B\kern-.05em{\sc i\kern-.025em b}\kern-.08em
    T\kern-.1667em\lower.7ex\hbox{E}\kern-.125emX}}
\begin{document}
\bstctlcite{IEEEexample:BSTcontrol}

\title{Reading the Mood: Emotion-Guided Book-to-Music Recommendation via CGANs and LLMs}

\author{\IEEEauthorblockN{Manousos Linardakis}
\IEEEauthorblockA{School of Electrical \& Computer Engineering\\
National Technical University of Athens\\
Zografou, 15780, Greece \\
manousoslinardakis@mail.ntua.gr}
\and
\IEEEauthorblockN{Georgios Alexandridis}
\IEEEauthorblockA{Department of Digital Industry Technologies\\
National \& Kapodistrian University of Athens\\
Evripos Complex, Psachna, 34400, Greece \\
gealexandri@uoa.gr}
}

\maketitle

\AddToShipoutPictureFG*{%
  \AtPageLowerLeft{%
    \put(\LenToUnit{0.5\paperwidth},\LenToUnit{1.5cm}){%
      \makebox[0pt][c]{\parbox{0.85\paperwidth}{\centering\tiny
        \copyright~2026 IEEE. Personal use of this material is permitted. Permission from IEEE must be obtained for all other uses, in any current or future media, including reprinting/republishing this material for advertising or promotional purposes, creating new collective works, for resale or redistribution to servers or lists, or reuse of any copyrighted component of this work in other works.}}}}}

\begin{abstract}
Background music that matches the mood of a text has been shown to make readers feel more immersed and improve their reading experience, motivating recommender systems that pair books with mood-matched music. In this direction, we present \emph{Sentiment Aware Generative Adversarial Network for Cross Domain Recommendation} (SAGA-CDR), a two-phase cross-domain recommendation framework that personalizes music suggestions \emph{and} emotionally aligns them with the book being read. In the first phase, transformer-based sentiment embeddings are constructed from user reviews and mapped across domains via a Conditional Generative Adversarial Network, whose mask-conditioned generator handles missing sentiment components and injects stochasticity for richer preference transfer. A compact rating neural network then fuses sentiment-specific interaction scores with a collaborative filtering prior to predict music ratings. In the second phase, large language models classify each book into a valence--arousal emotional quadrant, and candidate tracks are filtered to match that quadrant. Experiments on both the English Amazon and Chinese Douban datasets show that SAGA-CDR achieves the best rating prediction accuracy on Amazon (RMSE $0.98$) and the lowest RMSE on Douban ($0.91$), with ranking performance competitive with the strongest sentiment-aware baseline, even in cross-lingual settings.
\end{abstract}


\section{Introduction}
\label{sec:intro}

Recommender systems (RSs) operating within a single domain cannot leverage preferences expressed across different content types. Cross-domain recommendation (CDR) addresses this shortcoming by transferring knowledge from a source domain, where user activity is abundant, to a target domain where data are sparse. Most CDR methods rely on structural signals such as shared latent factors or user-item interaction patterns, ignoring the rich emotional information embedded in user-generated text.

In the books \& music case, emotional alignment is particularly well-motivated. Experimental evidence shows that emotionally congruent background music significantly enhances reading immersion and perceived text quality~\cite{hauck2023}. A RS for this scenario must therefore satisfy two simultaneous objectives: (i)~match individual user taste and (ii)~align with the emotional tone of the book being read. In order to address these issues, we propose  \emph{Sentiment Aware Generative Adversarial Network for Cross Domain Recommendation} (SAGA-CDR), a two-phase recommendation pipeline, whose main contributions are:
\begin{itemize}
	\item A formulation of emotion-guided book-to-music recommendation as a two-phase task, combining cross-domain preference transfer with valence–arousal congruence between the book and the recommended track.
	\item A cross-domain mapping in which the deterministic bridge used by prior sentiment-aware CDR methods is replaced by a conditional GAN with a presence mask over sentiment partitions and noise injection, addressing missing partitions and one-to-many transfer mechanisms. 
	\item An affective filtering stage in which LLMs assign books to VA quadrants, with the choice of LLM informed by a human study on cases where the models disagree. 
	\item An empirical evaluation on two public benchmarks, including cross-lingual transfer from sparse English data to dense Chinese data, with post-hoc SHAP and ICE analysis.
\end{itemize}

The rest of this paper is organized as follows; Section~\ref{sec:related} reviews related work and positions SAGA-CDR contribution, while Section~\ref{sec:methodoloy} lays out its core components. Section~\ref{sec:experiments} describes the experiments, Section~\ref{sec:results} presents the obtained results and Section~\ref{sec:explainability} analyzes the neural network prediction module in terms of the explainability of predictions. Finally, the paper concludes in Section~\ref{sec:conclusion}.

\section{Related Work}
\label{sec:related}

\subsection{Structural and Mapping Approaches}
\label{subsec:related:structural}

Traditional CDR methods often focus on structural signals, learning a bridge function between independently trained embedding spaces. EMCDR \cite{man2017} trains separate Matrix Factorization (MF) models per domain and fits a shared multi-layered perceptron (MLP) to translate source user embeddings to the target space, providing a cold-start profile for unseen users. PTUPCDR \cite{zhu2022} extends this approach with a meta-network that generates a personalised bridge function for each user, conditioned on a characteristic embedding derived from their transferable interests in the source-domain interaction history. More recently, HEAD \cite{choi2025based} leverages hyperbolic representations to model hierarchical user--item interactions while preserving power-law structures across domains. Meanwhile, CFAA \cite{liu2022collaborative} aligns BERT-based cross-domain user and item embeddings through Optimal Transport and graph-based attribution matching to mitigate domain discrepancy. However, these methods operate purely on interaction data and metadata, overlooking the rich affective signals embedded in review text. While some models, like RC-DFM \cite{fu2019deeply}, inject raw review text into Stacked Denoising Autoencoders to bridge latent spaces, sentiment is not extracted explicitly.

\subsection{Sentiment-Aware Approaches}
\label{subsec:related:sentiment}
 
To enrich representations beyond interaction patterns, a second line of work incorporates sentiment or emotion extracted from user reviews. CDR-SAFM~\cite{wang2020} partitions reviews into three polarity classes using a rating-based heuristic, applies Latent Dirichlet Allocation (LDA) to each class, and trains independent MLPs to map sentiment topic vectors across domains. STN-CDRS~\cite{taneja2024stn} addresses data sparsity by converting unrated text reviews into pseudo-ratings using sentiment analysis from a recurrent neural network. CRAS~\cite{zhang2024} uses VADER to discard negative reviews, retain only positive and neutral reviews for topic extraction, and train an improved CycleGAN to translate user preference matrices between domains. UTAM~\cite{azam2025utam} combines TextBlob and BERT for aspect-level sentiment, clustering users thematically and matching source-domain clusters to target-domain counterparts. Moving to richer affect representations, AAPAM~\cite{leung2020} extracts seven-dimensional emotion profiles via a Tweet Affective Classifier, identifying cross-domain user pairs by affective cosine similarity. ECDR~\cite{chakraverty2017review} builds six-dimensional emotion profiles derived from Parrott's hierarchical model, applying symmetric KL-divergence to identify emotionally similar target-domain users for collaborative filtering (CF). More recently, an affect-centric stance has been adopted by certain approaches~\cite{yilma2025}, by mapping music and visual art to a shared VA space, matching items by sentiment proximity using contrastive multimodal learning on audio, visual, and LLM-generated text features. Beyond sentiment, personality-aware RSs employ user traits to mitigate the cold start efect\cite{dhelim2022survey}, while responsible RSs prioritize the users' well-being and explainability over engagement \cite{kazienko2024toward}. SAGA-CDR shares both aims, using review sentiment instead of personality and explaining its predictions via SHAP/ICE and LLM justifications (Table~\ref{tab:examples}).

\subsection{SAGA-CDR}
\label{subsec:related:saga_cdr}

The literature on sentiment-aware CDR demonstrates a clear technological evolution from early lexicon-based tools and topic modeling, to deterministic MLPs, and finally to modern transformer-based architectures and multimodal learning. By utilizing a RoBERTa model fine-tuned on 124M tweets, we produce contextual 768-dimensional per-review embeddings that capture linguistic nuances missed by static LDA topics. Our main technical contribution is the replacement of deterministic bridges with a mask-conditioned stochastic Conditional GAN. Its explicit noise injection models the inherent one-to-many uncertainty in cross-domain preference transfer, while mask conditioning robustly handles users with missing sentiment partitions---practical data-quality issues often ignored by prior methods. 

\begin{figure}[htbp]
	\centerline{\includegraphics[width=\columnwidth]{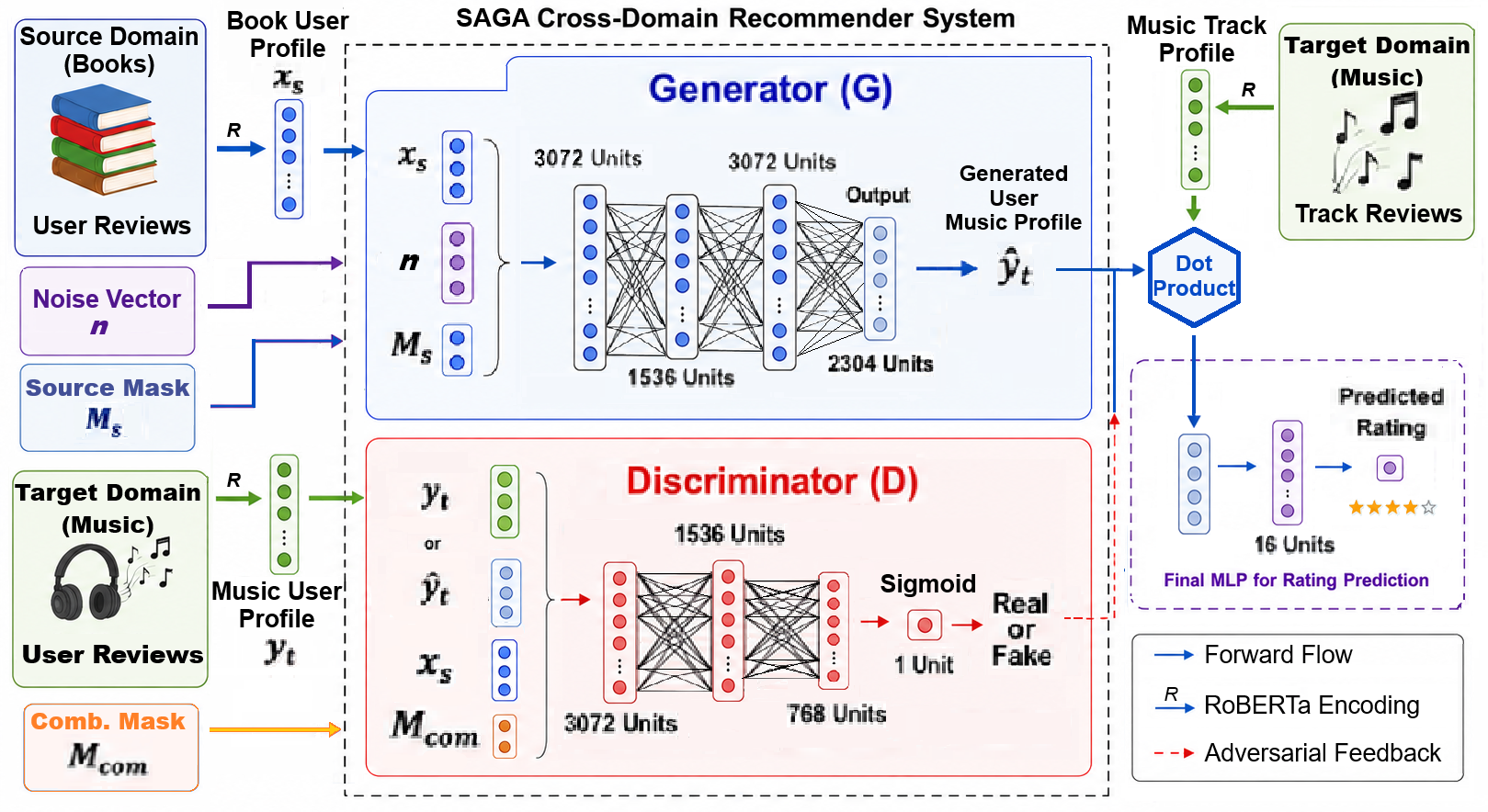}}
	\caption{The architectural diagram of the 1\textsuperscript{st} phase of SAGA-CDR} 
	\label{fig:framework}
\end{figure}

\section{Methodology}
\label{sec:methodoloy}

Fig.~\ref{fig:framework} provides a high-level overview of the 1\textsuperscript{st} phase of SAGA-CDR, outlining its two main components; (i) the CGAN used for domain transfer between the source and target domains and (ii) the neural network prediction module that performs the actual recommendation in the target domain. More formally, let $\mathcal{U}$ be a set of users with interactions in both a source domain $\mathcal{S}$ (e.g. books) and a target domain $\mathcal{T}$ (e.g. music). Each interaction is a quadruple $(u, i, r_{u,i}, d_{u,i})$ where $r_{u,i}$ is a numerical rating and $d_{u,i}$ a textual review. The goal is to predict ratings $\hat{r}_{u,i_t}$ for music items $i_t\in\mathcal{T}$ using book interactions as the primary signal.

\subsection{Sentiment-Aware Representations}
\label{subsec:method:sentiment}

For each review $d$ in the source domain, a RoBERTa-base model trained on 124\,M tweets and fine-tuned for sentiment classification via the TweetEval benchmark~\cite{loureiro2022} produces a contextual embedding $e_d\in\mathbb{R}^{768}$ (i.e. the \texttt{[CLS]} token representation) and a sentiment label $s_d\in\{\text{pos},\,\text{neu},\,\text{neg}\}$. For each entity $e$ (user or item), reviews are partitioned by sentiment class and mean-pooled within each class: $z^s_e = \frac{1}{|R^s_e|}\sum_{d\in R^s_e} e_d, s\in\{\text{pos},\,\text{neu},\,\text{neg}\}.$ The final representation is the concatenation $z_e = (z^{\text{pos}}_e,\, z^{\text{neu}}_e,\, z^{\text{neg}}_e) \in\mathbb{R}^{2304}$. Prior to training, all embeddings are z-score standardized using statistics computed exclusively on source-domain training data, placing all dimensions on a comparable scale while preventing information leakage.

\subsection{Cross-Domain Mapping via Conditional GAN}
\label{subsec:method:cgan}

The input to the GAN Generator $G$ (Fig.~\ref{fig:framework}, top middle) is comprised of (i)~the source domain embedding $z^s_u\in\mathbb{R}^{2304}$; (ii)~a noise vector $n\sim\mathcal{N}(0,I)$, $n\in\mathbb{R}^{100}$; and (iii)~a binary presence mask $m^s\in\{0,1\}^3$ indicating which sentiment components exist in the source domain. Batch normalization, the ReLU activation function and dropout are used in generator's hidden layers. 

The Discriminator $D$ (Fig.~\ref{fig:framework}, bottom middle) receives the concatenated source and target embeddings together with a combined mask $m^\text{comb} = m^s \vee m^t$ of both domains, outputting a real/fake probability. A key training asymmetry reflects inference conditions: $G$ is restricted to source-domain information only (mirroring cold-start deployment), while $D$ sees both domains, letting it learn what a plausible source–target pair looks like and provide stronger gradient feedback to $G$. Leaky ReLU ($\alpha{=}0.2$) and dropout ($p{=}0.3$) are used in discriminator's hidden layers.

The loss function of $D$ averages two BCE terms over real and generated pairs: $\mathcal{L}_D = -\frac{1}{2}\bigl( \mathbb{E}[\log D(z^s, z^{t}, m^{\text{comb}})] + \mathbb{E}[\log(1 - D(z^s, \hat{z}^{t}, m^{\text{comb}}))] \bigr)$.

The generator loss $\mathcal{L}_G = \mathcal{L}_\text{adv} + \lambda\,\mathcal{L}_\text{recon}$ consists of a non-saturating adversarial loss $\mathcal{L}_{\text{adv}} = -\mathbb{E}[\log D(z^s,
\hat{z}^t, m^{\text{comb}})]$ and a masked reconstruction term $\mathcal{L}_\text{recon} = \frac{1}{N}\sum_{i=1}^{N} \bigl\|(\hat{z}^t_i - z^t_i)\odot m^\text{comb}_i\bigr\|^2$. The mask applied to $\mathcal{L}_\text{recon}$ is deliberate. Components present in both domains are penalized normally; components present in only one domain teach the generator to produce a near-zero embedding where appropriate (since $m^{\text{comb}}$ retains them against a zero-filled target); and components absent in both domains are excluded entirely, as there is no ground truth to compare against.

\subsection{Rating Prediction Module} 
\label{subsec:method:rating_mlp}

The rating prediction module is a two-layered $4\rightarrow16\rightarrow1$ MLP (Fig.~\ref{fig:framework}, right) with ReLU activations, whose input vector is $x_{u,i} = (\text{score}^\text{pos},\, \text{score}^\text{neu},\, \text{score}^\text{neg},\, \text{base}(u,i))$, where each score is the dot product between the corresponding sentiment partition of the generated user embedding and the item embedding. The fourth feature is a bias-based CF prior $\text{base}(u,i) = \mu + b_u + b_i$, where $\mu$ is the global average rating in the target domain, $b_u$ is the \emph{user bias} (the difference between the user's mean rating in the source domain and the global source-domain average, capturing their general tendency to rate high or low) and $b_i$ is the \emph{item bias} (the difference between the item's mean rating in the target training set and $\mu$, reflecting its overall popularity--$b_i{=}0$ for cold-start items). Source-domain user averages are estimated from each user's available history at inference time; this does not constitute information leakage, as the source domain is treated as fully observed for both training and test users.

All four features are normalized using a quantile transformer fitted on the training set, mapping each value to its empirical quantile and applying the probit transform $\tilde{x} = \Phi^{-1}(p)$ (where $\Phi^{-1}(p)$ denotes the inverse CDF of the standard normal distribution), yielding approximately Gaussian marginals $\tilde{x} \sim \mathcal{N}(0,1)$, while preserving rank information. Quantile normalization is used because the raw features are strongly non-Gaussian (skewed, heavy-tailed, or sharply peaked), and it places them on a common scale while preserving rank.

The model is trained with a weighted combination of MSE and MAE: $\mathcal{L} = \alpha\,\mathcal{L}_\text{MSE} + (1-\alpha)\,\mathcal{L}_\text{MAE}, \text{with } \alpha = 0.7$.

We set $\alpha = 0.7$ to prioritize RMSE, since large mispredictions are more damaging than small frequent deviations.

\subsection{Emotion-Driven Recommendation}
\label{subsec:method:edr}

In emotion-driven recommendation (2\textsuperscript{nd} phase), the recommendation process is enhanced through emotional alignment between books and music. Book summaries are analyzed using LLMs to determine their dominant emotional tone within a VA framework. Music tracks are also categorized according to their emotional characteristics. The system then filters predicted recommendations by selecting music items that share the same emotional quadrant as books the user has enjoyed, thereby producing recommendations that are not only personalized but also emotionally consistent with the user's reading experience. The two phases are decoupled deliberately: quadrant classification runs offline (16.73 s per summary vs. 34.0 ms per forward pass), keeping online scoring lightweight and letting the filter's strictness change without retraining the mapping. The quadrant is shared by all readers of a book; personalization comes from Phase 1, as the filter acts on each user's own music candidate pool. 

Algorithm~\ref{alg:pipeline} summarizes the full pipeline. In Phase~1 (lines 1--16), Douban reviews are first translated to English, every review is encoded into a sentiment label and embedding, and these are mean-pooled into per-sentiment user and item representations. The CGAN is then trained to map source-domain user embeddings to the target domain, after which the rating module is trained on the quantile-normalized features. In Phase~2 (lines 17--25), each book summary is assigned a VA quadrant by Qwen3 offline; at recommendation time, the generator produces the user's music-domain profile, the rating module scores each candidate track, and only tracks with $\hat{r} \geq 4$ that match the book's quadrant are kept, ranked by predicted rating.

\begin{algorithm}[ht]
	\caption{Two-Phase Emotion-Guided CDR Framework}
	\footnotesize
	\begin{algorithmic}[1]
		\REQUIRE Source interactions $\mathcal{S}$, Target interactions $\mathcal{T}$, Book summaries $\mathcal{B}$, Tracks $\mathcal{M}$
		\ENSURE Target rating predictions $\hat{r}$, Affective candidate rankings $\mathcal{R}$
		\STATE \textbf{// Phase 1: Pre-processing \& Sentiment Extraction}
		\FOR{each Douban review $d \in \mathcal{T}$}
		    \STATE $d \leftarrow \text{Helsinki-NLP-Translate}(d)$ \COMMENT{zh $\rightarrow$ en}
		\ENDFOR
		\FOR{each review $d \in \mathcal{S} \cup \mathcal{T}$}
		    \STATE $s_d, e_d \leftarrow \text{TweetEval}(\text{RoBERTa})(d)$ 
		    \COMMENT{$s_d \in \{\text{pos, neu, neg}\}$}
		\ENDFOR
		\STATE Construct user and item sentiment arrays $z_u, z_i$ via mean-pooling
		\WHILE{CGAN not converged}
		    \STATE Sample batch $(z^s, z^t)$ and masks $m^{\text{comb}}$
		    \STATE Sample noise vector $n \sim \mathcal{N}(0, I)$
		    \STATE $\hat{z}^t \leftarrow \text{Generator}(z^s, n, m^s)$
		    \STATE Update Discriminator via $\mathcal{L}_D$
		    \STATE Update Generator via $\mathcal{L}_G$ 
		\ENDWHILE
		\STATE Train the rating module using Quantile-transformed features and target ratings via combined MSE/MAE loss ($\alpha = 0.7$)
		\STATE \textbf{// Phase 2: Affective Filtering \& Recommendation}
		\FOR{each book summary $b \in \mathcal{B}$}
		    \STATE $\mathcal{Q}_{book} \leftarrow \text{Qwen3\_Classification}(b)$
		\ENDFOR
		\FOR{each track $t \in \mathcal{M}$}
		    \STATE Generate music embedding for user via Generator
		    \STATE Predict rating $\hat{r}_{u,t}$ 
		\ENDFOR
		\STATE Filter candidates to match $\mathcal{Q}_{book}$ and sort by $\hat{r}_{u,t}$ descending
	\end{algorithmic}
	\label{alg:pipeline}
\end{algorithm}

\section{Experiments}
\label{sec:experiments}

All experiments were conducted on an ASUS Dual GeForce RTX 3060 GPU with 12\,GB of VRAM. The codebase, models and processed datasets for this paper are available in the accompanying repository.\footnote{\url{https://github.com/manouslinard/book-music-cdrs}} Average runtimes are 34.0 ms per prediction on Amazon and 6.3 ms on Douban, against 16.73 s per book summary for the offline Phase 2 classification (cf. Section~\ref{subsec:method:edr}).

\subsection{Datasets and Baselines}
\label{subsec:experiments:datasets-baselines}

\subsubsection{Amazon (Books $\to$ CDs \& Vinyl)}
Obtained from the Amazon\,2023 review corpus~\cite{hou2024}, using Books as the source domain and CDs \& Vinyl as the target. With only ${\approx}\,3.5$ music reviews per user, this benchmark represents a challenging and realistic sparse regime. Experiments are repeated across three independent seeds ($\{0, 42, 2026\}$) with 80k/20k/30k train/val/test users per seed; pairwise test-set user overlap is 3.2--3.4\%, confirming each run evaluates the model on a substantially distinct user subset.

\subsubsection{Douban (Books $\to$ Music)}
A Chinese-language benchmark from the Douban social network~\cite{zhu_douban2019}, with ${\approx}\,101.7$ music reviews per user---roughly 30$\times$ denser than Amazon. Evaluated via 5-fold cross-validation (1249/139/348 train/val/test users per fold) on seed$=42$. 

For both datasets, we retain only users with reviews in both domains, since CDR requires learning and assessing a user-level transfer function from source preferences to target behavior. Cold-start items unseen during training ($I_\text{cold} = I_\text{val/test} \setminus I_\text{train}$) are handled via TF-IDF category matching in the Amazon dataset. For each unseen category $c \notin C_\text{train}$, the most similar training category $\tilde{c}$ is identified via cosine similarity over TF-IDF vectors. Each cold-start item's category is replaced with $\tilde{c}$, and its embeddings are imputed with the mean of training items in that mapped category. For Douban (no category metadata), cold-start items receive the global mean training embedding. Test users' target-domain interactions serve only as ground truth: at inference the model sees exclusively the user's source-domain (book) profile, making the evaluation a strict target-domain cold-start setting.

\subsubsection{Baselines}

We compare SAGA-CDR against three CDR baselines, selected specifically to evaluate performance with and without textual signals. The first two are \textbf{EMCDR}~\cite{man2017}  and \textbf{PTUPCDR}~\cite{zhu2022}. Both use MF, with the former mapping embeddings through a deterministic MLP and the latter on a per-user meta-network. They fall under the structural and mapping category of systems (Section~\ref{subsec:related:structural}) and act as rating-only control groups. We intentionally include these text-agnostic models to assess the performance of methods that operate purely on ratings.

To ensure a fair comparison against models that do utilize rich review text, our third baseline is \textbf{CDR-SAFM}~\cite{wang2020}, which employs LDA topic features stratified by sentiment polarity, with three MLP regressors for cross-domain mapping. It is a sentiment-aware approach (Section~\ref{subsec:related:sentiment}), as it implements explicit, personalized sentiment-feature mapping. 

We chose these specific baselines, because more recent text-based methods use sentiment mainly as an intermediate step, to impute missing ratings as pseudo-ratings~\cite{taneja2024stn}, filter reviews~\cite{zhang2024}, or form user clusters~\cite{azam2025utam}, rather than as the representation mapped across domains. Furthermore, state-of-the-art affect models (e.g., \cite{yilma2025}) require multi-modal inputs, such as raw audio or visual features. Evaluating them in our setting would require modifying their formulation, so an observed gap could reflect our adaptation rather than the method itself. 

For EMCDR and PTUPCDR, we employ the official implementations \cite{zhu2022}, while CDR-SAFM is re-implemented from~\cite{wang2020}; all systems share the same train/validation/test user splits.

\subsubsection{Metrics}
 
We evaluate all systems on a set of standard offline RS metrics \cite{10.1145/3556536}, including \textbf{prediction accuracy} ones (\textit{RMSE} and \textit{MAE}) and \textbf{ranking} ones (Hit Ratio at $k$ -- \textit{HR@}$k$, Mean Average Precision at $k$ -- \textit{MAP@}$k$, Normalized Discounted Cumulative Gain -- \textit{NDCG@}$k$, and Mean Reciprocal Rank at $k$ -- \textit{MRR@}$k$) for various cut-off values ($k\in\{1,3,5,10\}$ -- higher is better), with an item deemed relevant if its true rating is at least 4.0. Users with no relevant test items receive a score of zero on all ranking metrics.

\subsubsection{Statistical Significance}

For the Amazon dataset ($n{=}3$ seeds), one-sided paired $t$-test $p$-values are reported comparing the SAGA-CDR against each baseline (significance threshold $\alpha{=}0.05$). The alternative hypothesis is that SAGA-CDR's mean is strictly better (lower for error metrics, higher for ranking metrics).

\subsection{Hyper-parameters}
\label{subsec:experiments:hyperparams}

Hyperparameters were selected using Copeland's method~\cite{saari1996copeland} across three random seeds ($0, 42$ and $2026$). Four CGAN configurations (Section \ref{subsec:method:cgan}) were evaluated over ($lr_G$, $lr_D$, $\lambda$): $(2 \times 10^{-4}, 2 \times 10^{-4}, 1.0)$, $(2 \times 10^{-4}, 2 \times 10^{-4}, 5.0)$, $(4 \times 10^{-4}, 1 \times 10^{-4}, 5.0)$ and $(4 \times 10^{-4}, 1 \times 10^{-4}, 10.0)$, with the third being selected. Selection was based on validation-set cosine similarity between generated and target embeddings. The asymmetric learning rates---with a higher rate for the generator---allow for more aggressive generator updates while keeping discriminator training stable. $\lambda = 5.0$ outperformed both extremes, indicating that neither suppressing the adversarial signal ($\lambda \geq 10$) nor deprioritizing reconstruction ($\lambda = 1.0$) is beneficial.

\begin{figure}[t]
  \centerline{\includegraphics[width=\columnwidth]{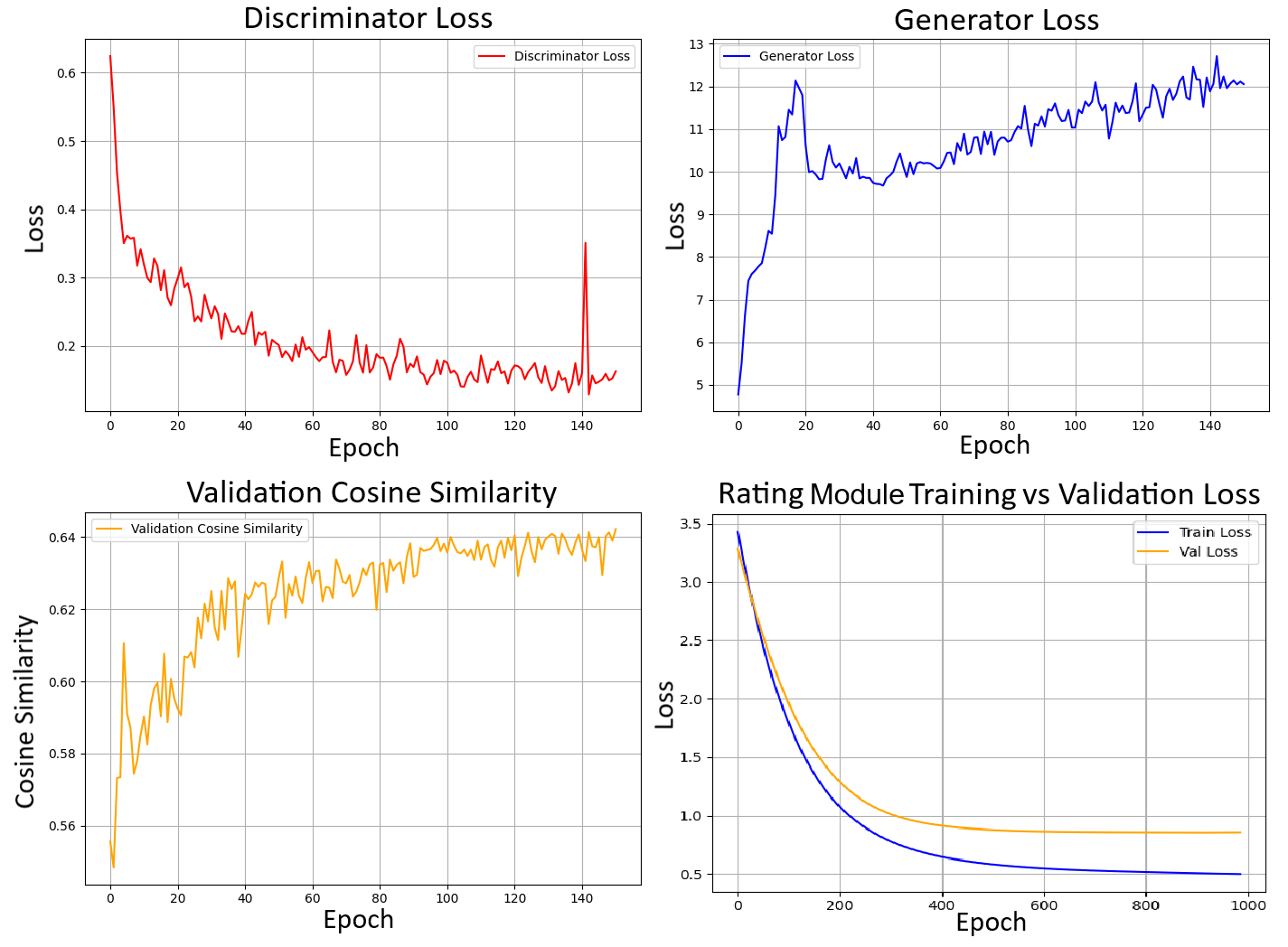}}
  \caption{SAGA-CDR training curves (in the Amazon dataset) after hyperparameter selection.}
  \label{fig:framework_curves}
\end{figure}

CGAN training used a batch size of 512 and a 200-epoch schedule with early stopping (patience 30), which terminated at approximately epoch 145. The training curves are displayed in Fig.~\ref{fig:framework_curves}. The discriminator loss decreases steadily but never approaches zero, so the discriminator does not saturate and the generator continues to receive informative gradients throughout training. Correspondingly, the generator's loss rises and then stabilizes. The validation cosine similarity between generated and target embeddings rises steadily before plateauing at about 0.64. Training does not diverge: the only irregularity is a single spike in the discriminator loss near epoch 141, which is absorbed within one epoch. The downstream rating network, trained on the generated embeddings with a 1000-epoch budget, also trains smoothly. Its training and validation losses (Fig.~\ref{fig:framework_curves}, bottom right) decrease together, with the validation loss flattening at about 0.86 by roughly epoch 600 and not rising afterwards, showing no sign of overfitting.

\subsection{Emotion-Driven Recommendation (Phase~2)} 
\label{subsec:experiments:phase_two}

\subsubsection{Music dataset construction.}
The MERGE Audio Dataset~\cite{louro2024} provides VA-annotated tracks, which are further discretised into four quadrants: (i) Q1 (energetic/positive: Pop, R\&B), (ii) Q2 (energetic/negative: Metal, Punk), (iii) Q3 (calm/negative: Soul, Blues), (iv) Q4 (calm/positive: Jazz, Bossa Nova)~\cite{russell1980}. To connect it to the Amazon ecosystem, both datasets are normalized (lowercased, punctuation stripped) and joined on the resulting title field to produce a candidate metadata table $M_\text{meta}$. To guard against spurious title matches across different artists, a strict artist-name validation filter is applied: a match is retained only if the MERGE artist name appears in at least one Amazon text field (product description, review text, store name, or title). Of the 3554 MERGE tracks, 1372 match an Amazon title and 365 survive artist validation, forming $M_\text{final}$, which maps each song's audio file and emotional quadrant to its Amazon ASIN.

\subsubsection{Book dataset construction}

The CMU Book Summary Dataset~\cite{bamman2013} provides Wikipedia plot summaries. These are aligned with the Amazon Books catalogue via the same normalized title-based join as $M_\text{meta}$, yielding $B_\text{meta}$. An analogous author-name validation filter retains a match only if the CMU author name appears in the corresponding Amazon review text or item title. Of the 16277 CMU titles, 11704 match an Amazon title and 6818 survive author validation, producing $B_\text{final}$, which associates each book's title and summary with its Amazon ASIN for later LLM classification.

\begin{figure}[htbp]
	\centering
	\resizebox{0.96\columnwidth}{!}{%
		\fbox{\parbox{\columnwidth}{%
			\small
			\textbf{System framing:} Act as a musical curator for literature. Choose
			one of four musical quadrants as the single, overarching soundtrack for a book.
			Imagine the user will listen for the \emph{entire} reading experience. Your choice
			must represent the book's dominant, persistent mood.\\[2pt]
			\textbf{Q1} (High Arousal, High Valence): uplifting, triumphant, fast-paced.
			Genres: Dance-Rock, R\&B, Pop, Synth-Pop, Indie Pop.\\
			\textbf{Q2} (High Arousal, Low Valence): tense, aggressive, dark, chaotic.
			Genres: Heavy Metal, Punk, Grunge, Industrial.\\
			\textbf{Q3} (Low Arousal, Low Valence): melancholic, somber, deeply reflective.
			Genres: Soul, Blues, Adult Contemporary, Neo-Soul.\\
			\textbf{Q4} (Low Arousal, High Valence): serene, peaceful, gentle, relaxing.
			Genres: Jazz, Easy Listening, Bossa Nova, Soft Rock.\\[2pt]
			\textbf{Output format:} Start with Q1, Q2, Q3, or Q4, then colon, then reasoning.\\
			\emph{Example:} Q2: The story's constant suspense and dark themes align with
			tense, dramatic music.\\[2pt]
			\textbf{Book:} \{title\} \quad \textbf{Author:} \{author\} \quad
			\textbf{Summary:} \{summary\}
		}}%
	}
	\caption{LLM prompt template for emotional quadrant classification. All three
	models receive the same prompt; only the book metadata varies per instance.}
	\label{fig:prompt}
\end{figure}

\subsubsection{LLM book classification}

Three open-source LLMs (LLaMA\,3.1\,8B~\cite{llama3}, Qwen3\,8B~\cite{qwen3}, and DeepSeek-R1\,8B~\cite{deepseek}) are run locally via Ollama (temperature\,$=$\,0, fixed seed $=42$) to classify each book into a VA quadrant. Each model receives the same structured prompt (Fig.~\ref{fig:prompt}): it is framed as a \emph{musical curator for literature} asked to select the quadrant representing the book's \emph{dominant, persistent mood} as if the music will accompany the full reading experience. The prompt defines each quadrant with VA coordinates, a qualitative description, and representative genres, then requires a structured \texttt{Q\{1--4\}: <reasoning>} output. This structured output enables reliable automated parsing, while the natural-language justification lets annotators audit the models' reasoning in disputed cases and could be surfaced to end-users as an explanation.

Majority vote~\cite{trad2024ensemble} resolves cases where at least two models agree; of the 129 books, 53 received unanimous assignments, 70 were resolved by a 2-to-1 majority, and 6 were fully disputed (all three models disagreeing) and forwarded to human evaluation. These 129 are the subset of $B_\text{final}$ rated above 4 stars by evaluation users, since a soundtrack is only meaningful for a book the reader actually enjoyed. A form was generated from the LLM output and for each disputed book, annotators were shown the title, an author-curated one-paragraph summary sourced from Wikipedia and Goodreads (to avoid LLM-influenced wording), and the reasoning provided by each disagreeing model for its proposed quadrant. Reviewers were instructed to select the quadrant best representing the book's \emph{dominant, persistent} mood---mirroring the framing used in the LLM prompt.

The study involved 20 gender-balanced participants (ages 18–44, predominantly holding a BSc or higher), whose ML experience ranged from none (25\%) to advanced (10\%), intentionally mixing typical end-users with practitioners to improve representativeness.

LLM quality was assessed via a combined score: $\text{final}(k) = (1{-}w)\,\hat{v}(k) + w,\text{ } \hat{h}(k), \quad w = 0.6$, where $\hat{v}(k)$ is the min-max normalised weighted inter-model agreement score (unanimous agreement rewarded more than a 2-to-1 split), and $\hat{h}(k)$ is the normalised human evaluation score (number of disputed books for which $k$'s quadrant was selected by the majority of annotators). The weight $w{=}0.6$ reflects the greater reliability of human judgement over inter-model agreement.
 
DeepSeek-R1 and Qwen3 scored comparably on inter-model agreement (54 and 55 points respectively), while LLaMA\,3.1 scored considerably lower (31), suggesting it more frequently proposed the minority quadrant. On the 6 disputed cases, Qwen3 was preferred by human annotators in 3 books, compared to 2 for LLaMA\,3.1 and only 1 for DeepSeek-R1. Qwen3 achieves a final score of 1.00 vs. 0.38 for DeepSeek-R1 and 0.30 for LLaMA 3.1; since it leads on both components, it ranks first for any $w \in [0, 1]$, although six disputed books remain a small sample.

\begin{table}[htbp]
	\caption{Sample reading soundtrack recommendations. Q: emotional quadrant.}
	\begin{center}
	\resizebox{\columnwidth}{!}{%
		\begin{tabular}{p{2cm}p{1.55cm}cp{3.05cm}c}
			\hline
			\textbf{Book} & \textbf{Music} & \textbf{Q} & \textbf{LLM reasoning} (summary) & \textbf{Rating} \\
			\hline
			\textit{The Heroin Diaries} (Nikki Sixx)
			  & ``Sing Sing Death House'' -- The Distillers & Q2 & Addiction, despair, turmoil, and self-destruction. & 4.8 \\
			\textit{Foley Is Good} (Mick Foley)
			  & ``I Can't Stop'' -- Al Green & Q1 & Celebratory, energetic, positive, and triumphant mood. & 4.75 \\
			\textit{Franny and Zooey} (J.D. Salinger)
			  & ``Is This Desire?'' -- PJ Harvey & Q3 & Introspective, somber, emotionally fragile, and existential. & 4.7 \\
			\textit{The Seven Storey Mountain} (Thomas Merton)
			  & ``Prelude to a Kiss'' -- Duke Ellington & Q4 & Reflective spiritual journey, transformation, and inner peace. & 4.6 \\
			\hline
		\end{tabular}
	}
	\end{center}
	\label{tab:examples}
\end{table}

\subsubsection{Final recommendation}

For a book rated more than $4$ stars in the Amazon dataset, Qwen3 assigns its quadrant. The recommended track candidate pool ($\hat{r}\geq 4.0$) is filtered to the matching quadrant and ranked by predicted score. Table~\ref{tab:examples} shows the reading soundtrack recommendations produced by the full pipeline. Since a recommendation must satisfy both criteria, the trade-off appears as coverage rather than quality: at least one candidate survives for 46 of 139 (user, book) pairs (33.1\%), and the mean predicted rating among surviving pairs is statistically unchanged (4.591 before and after filtering, paired t-test p = 0.85), since the filter removes candidates outright rather than reordering them. Coverage is bounded by the catalogue (365 tracks, 129 books), not the method.

\section{Results}
\label{sec:results}

Table~\ref{tab:amazon_pred} shows that SAGA-CDR achieves the best rating prediction performance for the Amazon dataset by a substantial and statistically significant margin, on both prediction accuracy metrics. The RMSE of 0.9806 and MAE of 0.6097 represent a 12.5\% relative RMSE improvement over CDR-SAFM, the closest sentiment-aware competitor. 

\begin{table}[htbp]
	\caption{Amazon rating prediction (mean\,$\pm$\,std, $n{=}3$ seeds) \\ \scriptsize $\dagger$: significance against SAGA-CDR ($\alpha{=}0.05$) -- Lower is better}
    \label{tab:amazon_pred}
	\begin{center}
	\resizebox{\columnwidth}{!}{%
		\begin{tabular}{l|cc|cc}
		\hline
		\textbf{Method} & \textbf{RMSE}\,$\downarrow$ & $p$ & \textbf{MAE}\,$\downarrow$ & $p$ \\
		\hline
		\textbf{SAGA-CDR} & \textbf{0.9806\,$\pm$\,0.0078} & ---
		& \textbf{0.6097\,$\pm$\,0.0145} & --- \\
		CDR-SAFM & 1.1204\,$\pm$\,0.0018 & 0.0006$^\dagger$
		& 0.6674\,$\pm$\,0.0069 & 0.0145$^\dagger$ \\
		EMCDR    & 4.6179\,$\pm$\,0.0129 & 0.0000$^\dagger$
		& 4.5115\,$\pm$\,0.0129 & 0.0000$^\dagger$ \\
		PTUPCDR  & 4.6742\,$\pm$\,0.0084 & 0.0000$^\dagger$
		& 4.5011\,$\pm$\,0.0147 & 0.0000$^\dagger$ \\
		\hline
		\end{tabular}
	}
	\end{center}
\end{table}

\begin{table*}[htbp]
	\caption{Amazon ranking metrics (mean\,$\pm$\,std, $n{=}3$ seeds). $\dagger$: significant. Higher is better.}
	\begin{center}
    \renewcommand{\arraystretch}{0.88}
		\begin{tabular}{clcccc}
			\hline
			$\mathbf{k}$ & \textbf{Method} & \textbf{MAP}\,$\uparrow$ & \textbf{NDCG}\,$\uparrow$ & \textbf{MRR}\,$\uparrow$ & \textbf{HR}\,$\uparrow$ \\
			\hline
			\multirow{4}{*}{1}
			 & \textbf{SAGA-CDR} & \textbf{.8907\,$\pm$.0005} & \textbf{.8907\,$\pm$.0005} & \textbf{.8907\,$\pm$.0005} & \textbf{.8907\,$\pm$.0005} \\
			 & SAFM  & .8902\,$\pm$.0019 ($p{=}.2955$) & .8902\,$\pm$.0019 ($p{=}.2955$) & .8902\,$\pm$.0019 ($p{=}.2955$) & .8902\,$\pm$.0019 ($p{=}.2955$) \\
			 & EMCDR & .8859\,$\pm$.0007 ($p{=}.0056^\dagger$) & .8859\,$\pm$.0007 ($p{=}.0056^\dagger$) & .8859\,$\pm$.0007 ($p{=}.0056^\dagger$) & .8859\,$\pm$.0007 ($p{=}.0056^\dagger$) \\
			 & PTUPCDR  & .8861\,$\pm$.0012 ($p{=}.0075^\dagger$) & .8861\,$\pm$.0012 ($p{=}.0075^\dagger$) & .8861\,$\pm$.0012 ($p{=}.0075^\dagger$) & .8861\,$\pm$.0012 ($p{=}.0075^\dagger$) \\
			\hline
			\multirow{4}{*}{3}
			 & \textbf{SAGA-CDR} & \textbf{.8946\,$\pm$.0011} & \textbf{.9026\,$\pm$.0014} & \textbf{.9075\,$\pm$.0012} & \textbf{.9254\,$\pm$.0020} \\
			 & SAFM  & .8947\,$\pm$.0019 ($p{=}.5243$) & .9027\,$\pm$.0020 ($p{=}.5561$) & .9071\,$\pm$.0020 ($p{=}.2732$) & .9255\,$\pm$.0021 ($p{=}.7887$) \\
			 & EMCDR & .8912\,$\pm$.0010 ($p{=}.0022^\dagger$) & .9000\,$\pm$.0011 ($p{=}.0051^\dagger$) & .9049\,$\pm$.0008 ($p{=}.0085^\dagger$) & .9255\,$\pm$.0019 ($p{=}.7100$) \\
			 & PTUPCDR  & .8910\,$\pm$.0015 ($p{=}.0179^\dagger$) & .9000\,$\pm$.0014 ($p{=}.0233^\dagger$) & .9049\,$\pm$.0012 ($p{=}.0116^\dagger$) & .9252\,$\pm$.0019 ($p{=}.2222$) \\
			\hline
			\multirow{4}{*}{5}
			 & \textbf{SAGA-CDR} & \textbf{.8973\,$\pm$.0014} & \textbf{.9063\,$\pm$.0016} & \textbf{.9078\,$\pm$.0012} & \textbf{.9267\,$\pm$.0020} \\
			 & SAFM  & .8976\,$\pm$.0019 ($p{=}.7192$) & .9064\,$\pm$.0020 ($p{=}.5851$) & .9074\,$\pm$.0020 ($p{=}.2572$) & .9268\,$\pm$.0021 ($p{=}.7887$) \\
			 & EMCDR & .8943\,$\pm$.0014 ($p{=}.0025^\dagger$) & .9041\,$\pm$.0015 ($p{=}.0041^\dagger$) & .9052\,$\pm$.0008 ($p{=}.0060^\dagger$) & .9267\,$\pm$.0021 ($p{=}.6291$) \\
			 & PTUPCDR  & .8942\,$\pm$.0016 ($p{=}.0139^\dagger$) & .9041\,$\pm$.0016 ($p{=}.0162^\dagger$) & .9053\,$\pm$.0012 ($p{=}.0101^\dagger$) & .9267\,$\pm$.0021 ($p{=}.5000$) \\
			\hline
			\multirow{4}{*}{10}
			 & \textbf{SAGA-CDR} & \textbf{.9004\,$\pm$.0012} & \textbf{.9095\,$\pm$.0015} & \textbf{.9078\,$\pm$.0012} & \textbf{.9270\,$\pm$.0021} \\
			 & SAFM  & .9005\,$\pm$.0018 ($p{=}.5987$) & .9095\,$\pm$.0019 ($p{=}.5000$) & .9074\,$\pm$.0020 ($p{=}.2278$) & .9270\,$\pm$.0020 ($p{=}.5000$) \\
			 & EMCDR & .8975\,$\pm$.0015 ($p{=}.0035^\dagger$) & .9075\,$\pm$.0016 ($p{=}.0025^\dagger$) & .9052\,$\pm$.0008 ($p{=}.0082^\dagger$) & .9270\,$\pm$.0020 ($p{=}.7887$) \\
			 & PTUPCDR  & .8976\,$\pm$.0016 ($p{=}.0137^\dagger$) & .9075\,$\pm$.0016 ($p{=}.0113^\dagger$) & .9053\,$\pm$.0012 ($p{=}.0117^\dagger$) & .9269\,$\pm$.0020 ($p{=}.0918$) \\
			\hline
		\end{tabular}
	\end{center}
	\label{tab:amazon_rank}
\end{table*}

Table~\ref{tab:amazon_rank} draws an equally clear picture for the rank accuracy metrics on the Amazon dataset. At $k{=}1$---the strictest test of whether the single best recommendation is immediately relevant---SAGA-CDR significantly outperforms both EMCDR \& PTUPCDR. This advantage is sustained and is also statistically significant across every cut-off and every precision-oriented metric (MAP, NDCG, MRR) for these two baselines. It is important to note that EMCDR and PTUPCDR still have high ranking metrics in absolute terms (above 88\% at $k{=}1$). Combined with their rating errors (Table~\ref{tab:amazon_pred}), however, it can be concluded that MF-based methods fail to perform as well in sparse regimes (${\approx}\,3.5$ reviews per user) as review-based methods, even when they order items reasonably.

Against CDR-SAFM, SAGA-CDR ties on all ranking metrics. In our setting, however, rating accuracy carries more weight. Phase 2 forms each user's candidate pool with an absolute threshold ($\hat{r} \geq 4.0$), so the specific tracks to be recommended depend on predictions being correct on the rating scale, a step that ranking metrics computed on Phase 1 do not capture. In rating prediction, SAGA-CDR clearly leads (RMSE = 0.98 vs. 1.12), making it the only method to pair the best prediction accuracy with tied-best ranking. For HR@$k$ (where $k>1$), all methods are statistically tied at every cut-off, suggesting that the main distinction is not whether a relevant item appears in the top-$k$ list, but how highly relevant items are ranked within that list.

Prior to analyzing the Douban results, an asymmetry in the setup should be noted: SAGA-CDR operates under a more demanding transfer regime, with the CGAN having been pre-trained on sparse English Amazon data - since the CGAN requires a larger paired cross-domain signal than Douban alone provides - and only the rating module having been fine-tuned on Douban. Once learned, the mapping transfers effectively even across language and density shifts (Table~\ref{tab:douban}).

\begin{table*}[htbp]
	\caption{Douban results (5-fold CV averages). Bold: best; underline: second best.}
	\begin{center}
	\resizebox{\textwidth}{!}{%
		\begin{tabular}{ll|cc|cccc|cccc|cccc|cccc}
			\hline
			& & \multicolumn{2}{|c|}{Pred.\,$\downarrow$} & \multicolumn{4}{|c|}{$k{=}1$\,$\uparrow$} & \multicolumn{4}{|c|}{$k{=}3$\,$\uparrow$} & \multicolumn{4}{|c|}{$k{=}5$\,$\uparrow$} & \multicolumn{4}{|c}{$k{=}10$\,$\uparrow$} \\
			\cline{3-4}\cline{5-8}\cline{9-12}\cline{13-16}\cline{17-20}
			& Method & RMSE & MAE & MAP & NDCG & HR & MRR & MAP & NDCG & HR & MRR & MAP & NDCG & HR & MRR & MAP & NDCG & HR & MRR \\
			\hline
			\multirow{3}{*}{\rotatebox{90}{\tiny pre-tr.}}
			 & SAFM    & 0.9169 & \textbf{0.6925} & 0.9118 & 0.9118 & 0.9118 & 0.9118 & 0.8848 & 0.9085 & 0.9775 & 0.9433 & 0.8769 & 0.9098 & \textbf{0.9839} & 0.9447 & 0.8704 & 0.9121 & \textbf{0.9839} & 0.9447 \\
			 & EMCDR   & 4.0819 & 3.9763 & 0.8341 & 0.8341 & 0.8341 & 0.8341 & 0.7963 & 0.8371 & 0.9637 & 0.8953 & 0.7850 & 0.8411 & 0.9798 & 0.8990 & 0.7790 & 0.8487 & 0.9833 & 0.8995 \\
			 & PTUPCDR    & 4.0793 & 3.9735 & 0.8272 & 0.8272 & 0.8272 & 0.8272 & 0.7877 & 0.8306 & 0.9672 & 0.8924 & 0.7785 & 0.8357 & 0.9781 & 0.8948 & 0.7742 & 0.8444 & 0.9833 & 0.8956 \\
			\hline
			\multirow{3}{*}{\rotatebox{90}{\tiny no-pre.}}
			 & SAFM    & \underline{0.9151} & \underline{0.6928} & 0.9084 & 0.9084 & 0.9084 & 0.9084 & 0.8866 & 0.9097 & 0.9781 & 0.9417 & 0.8775 & 0.9097 & \underline{0.9833} & 0.9429 & \textbf{0.8720} & \textbf{0.9131} & \textbf{0.9839} & 0.9430 \\
			 & EMCDR   & 3.5591 & 3.2499 & 0.8750 & 0.8750 & 0.8750 & 0.8750 & 0.8426 & 0.8758 & 0.9764 & 0.9236 & 0.8346 & 0.8792 & 0.9816 & 0.9248 & 0.8258 & 0.8820 & 0.9833 & 0.9251 \\
			 & PTUPCDR    & 3.8265 & 2.9725 & 0.8710 & 0.8710 & 0.8710 & 0.8710 & 0.8324 & 0.8670 & 0.9718 & 0.9184 & 0.8241 & 0.8713 & 0.9781 & 0.9210 & 0.8141 & 0.8741 & 0.9833 & 0.9211 \\
			\hline
			\multirow{3}{*}{\rotatebox{90}{\tiny SAGA-CDR}}
			 & notrans & 1.0155 & 0.7862 & \underline{0.9130} & \underline{0.9130} & \underline{0.9130} & \underline{0.9130} & 0.8859 & 0.9095 & \underline{0.9792} & \underline{0.9442} & 0.8792 & 0.9114 & 0.9827 & \underline{0.9450} & 0.8705 & 0.9124 & \textbf{0.9839} & \underline{0.9452} \\
			 & multi   & 0.9267 & 0.7525 & 0.9078 & 0.9078 & 0.9078 & 0.9078 & \underline{0.8914} & \underline{0.9135} & \underline{0.9792} & 0.9421 & \underline{0.8808} & \underline{0.9118} & 0.9821 & 0.9428 & \underline{0.8714} & 0.9123 & \textbf{0.9839} & 0.9430 \\
			 & \textbf{trans.} & \textbf{0.9057} & 0.7284 & \textbf{0.9165} & \textbf{0.9165} & \textbf{0.9165} & \textbf{0.9165} & \textbf{0.8915} & \textbf{0.9143} & \textbf{0.9810} & \textbf{0.9471} & \textbf{0.8810} & \textbf{0.9129} & \underline{0.9833} & \textbf{0.9477} & 0.8695 & \underline{0.9126} & \textbf{0.9839} & \textbf{0.9477} \\
			\hline
		\end{tabular}%
	}
	\end{center}
	\label{tab:douban}
\end{table*}

The competing baselines (CDR-SAFM, EMCDR, PTUPCDR) are evaluated both with and without Amazon pre-training, denoted in Table \ref{tab:douban} as \emph{pre-tr.} and \emph{no-pre.}, respectively, allowing a direct assessment of how much each baseline benefits from cross-dataset transfer versus training on Douban alone. SAGA-CDR is evaluated in three configurations, reflecting different strategies for handling the Chinese review text: (i) \emph{notrans} (no translation; Chinese-only RoBERTa), (ii) \emph{multi} (no translation; multilingual XLM-RoBERTa), and (iii) \emph{trans.} (reviews machine-translated to English via Helsinki-NLP~\cite{helsinki_translator}, then encoded with the same English-pretrained sentiment model used on the Amazon dataset). These models perform the initial sentiment recognition and user-review embedding generation, after which the resulting representations are passed to the remaining stages of SAGA-CDR.

Paired one-sided $t$-tests across the 5 CV folds, with Holm--Bonferroni correction per baseline over the 18 reported metrics, confirm SAGA-CDR's advantage over EMCDR and PTUPCDR on both prediction-accuracy metrics ($p<0.0001$ for all variants) and on MAP and NDCG at nearly every cut-off. Against CDR-SAFM, SAGA-CDR has lower RMSE and matches it on every ranking metric; CDR-SAFM leads only on MAE (p = 0.008) and only under cross-lingual transfer: on Amazon, SAGA-CDR leads on MAE as well (p = 0.0145).

\subsubsection{Prediction Accuracy}

Despite operating under the most demanding experimental conditions - cross-lingual transfer from a sparse English corpus to a dense Chinese one - SAGA-CDR attains the lowest RMSE ($0.9057$) of all methods evaluated on Douban, outperforming even baselines trained from scratch directly on the target data, while maintaining a competitive MAE of $0.7284$. The fact that SAGA-CDR, a model carrying cross-domain mappings from English Amazon reviews, matches or even exceeds target-trained competitors, constitutes the strongest evidence so far that the learned sentiment representations encode preference structure surviving both domain and language shift at the same time.

\subsubsection{Ranking Performance}

SAGA-CDR leads all methods on precision-sensitive ranking metrics at $k\in\{1,3,5\}$, taking the top position across MAP, NDCG, and MRR at these cut-offs, while holding the second-best position on HR@$5$ and remaining competitive at $k{=}10$ for most of the ranking metrics. Attaining this under cross-lingual transfer, where the closest competitor is trained directly on Douban, reflects the quality of the learned cross-domain user representations and their ability to distinguish highly relevant items from the rest of the candidate pool.
 
\subsubsection{Effect of Language Strategy}

The three variants reveal a clear hierarchy: translating reviews to English prior to encoding (\emph{SAGA-CDR}) is the strongest overall configuration, followed by multilingual encoding (\emph{multi}), with the monolingual Chinese model (\emph{notrans}) last, particularly on prediction accuracy. This ordering reflects representational alignment: the full training stack is anchored in English pretraining, so keeping review text in English eliminates cross-lingual mismatch. The multilingual model is a practical alternative when translation is unavailable, at a cost to prediction accuracy.

\subsection*{Ablation}

To isolate the contribution of each CGAN component, we retrain SAGA-CDR on Amazon with one component removed at a time, keeping the sentiment representations, rating module, hyper-parameters, seeds and splits unchanged, and test each variant against the full model with the same one-sided paired t-tests as in the Amazon experiments. Removing the adversarial term has the largest effect on rating prediction, raising RMSE from 0.981 to 1.002 and MAE from 0.610 to 0.640 (MAE: p = 0.013): without it, the generator is trained only to reconstruct target embeddings and loses the discriminator's signal about which source–target pairs are plausible. Removing noise injection, the presence mask or the reconstruction loss slightly lowers RMSE but raises MAE (0.621–0.636); neither change is statistically significant. On ranking metrics, all four variants exhibit worse performance, significantly below the full model on MAP, NDCG and MRR at $k{=}1$ and $k{=}10$. Table~\ref{tab:ablation} showcases these prediction and ranking metrics of the ablation study. Every variant still outperforms CDR-SAFM on RMSE and MAE, but only the full model matches it on ranking metrics. Therefore, all components of the full model play a meaningful role in achieving the overall system performance.

\begin{table}[t]
	\centering
	\caption{CGAN ablation on Amazon (mean over $n$=3 seeds). @1: MAP, NDCG and MRR. $\dagger$: significantly worse than full ($\alpha$=0.05).}
	\label{tab:ablation}
	\footnotesize
	\setlength{\tabcolsep}{4pt}
	\begin{tabular}{@{}lcccccc@{}}
		\toprule
		 & \multicolumn{2}{c}{Rating $\downarrow$} & \multicolumn{4}{c}{Ranking $\uparrow$} \\
		\cmidrule(lr){2-3}\cmidrule(l){4-7}
		Variant & RMSE & MAE & @1 & MAP@10 & NDCG@10 & MRR@10 \\
		\midrule
		Full model                   & 0.981 & \textbf{0.610}   & \textbf{.8907}  & \textbf{.9004}  & \textbf{.9095}  & \textbf{.9078} \\
		$-$ noise                    & 0.973 & 0.636            & .8857$^\dagger$ & .8979$^\dagger$ & .9075$^\dagger$ & .9049$^\dagger$ \\
		$-$ mask                     & 0.974 & 0.621            & .8857$^\dagger$ & .8979$^\dagger$ & .9075$^\dagger$ & .9049$^\dagger$ \\
		$-\,\mathcal{L}_\mathrm{adv}$   & 1.002 & 0.640$^\dagger$ & .8858$^\dagger$ & .8979$^\dagger$ & .9076$^\dagger$ & .9049$^\dagger$ \\
		$-\,\mathcal{L}_\mathrm{recon}$ & 0.973 & 0.625            & .8857$^\dagger$ & .8979$^\dagger$ & .9075$^\dagger$ & .9049$^\dagger$ \\
		\bottomrule
	\end{tabular}
\end{table}

\section{Model Explainability (XAI)}
\label{sec:explainability}

To evaluate the internal decision-making process of the rating prediction module, we apply SHAP~\cite{lundberg2017} and ICE~\cite{goldstein2015peeking} plots. The analyses reveal an interpretable feature-importance hierarchy: $\mathit{base\_term} \gg \mathit{score}_{pos} > \mathit{score}_{neg} \gg \mathit{score}_{neu}$

\begin{figure}[htbp]
	\centerline{\includegraphics[width=\columnwidth]{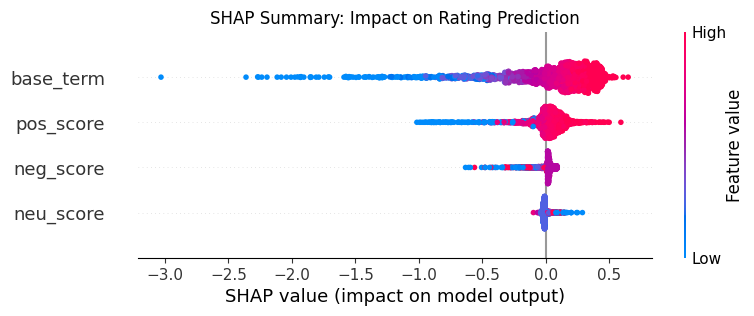}}
	\centerline{\includegraphics[width=\columnwidth]{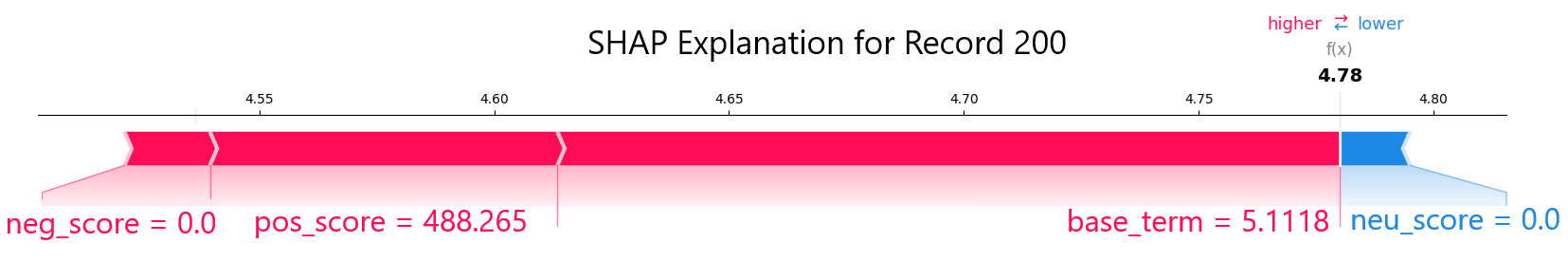}}
	\caption{SHAP explanations on Amazon dataset. Top: summary plot showing feature impact across all instances (red: high values, blue: low values). Bottom: force plot for record 200; red features increase the prediction, while blue (neu\_score) decreases it.} 
	\label{fig:xai}
\end{figure}

At the global level, the SHAP summary plot (Fig.~\ref{fig:xai}, top panel) shows that the CF prior ($\mathit{base\_term}$)  serves as the model's primary rating anchor. Its impact is asymmetric: low baseline values strongly penalize predictions, whereas high values yield only modest positive contributions. The $\mathit{score}_{pos}$ feature ranks second and displays a monotonically increasing positive influence, confirming that stronger positive sentiment reliably raises predicted ratings. 

The $\mathit{score}_{neg}$ feature shows a concentrated SHAP distribution centered around zero: negative sentiment acts as a secondary, context-dependent signal rather than a monotonic rating suppressor. Finally, $\mathit{score}_{neu}$ contributes the least, with SHAP values tightly clustered around zero, indicating that neutral sentiment carries limited discriminative power, reflecting its inherent ambiguity in recommender settings.

Fig.~\ref{fig:xai} (bottom panel) shows a local force plot for a representative instance (record $200$): \texttt{base\_term} ($= 5.112$) and \texttt{pos\_score} $= 488.265$ push the prediction upward, consistent with the global analysis. Notably, although \texttt{neg\_score} and \texttt{neu\_score} are both 0.0, they act in opposite directions; the absence of negative sentiment is an active positive signal, while zero neutral reviews introduces a mild downward correction to the final value of $4.78$.

\begin{figure}[htbp]
	\centerline{\includegraphics[width=\columnwidth]{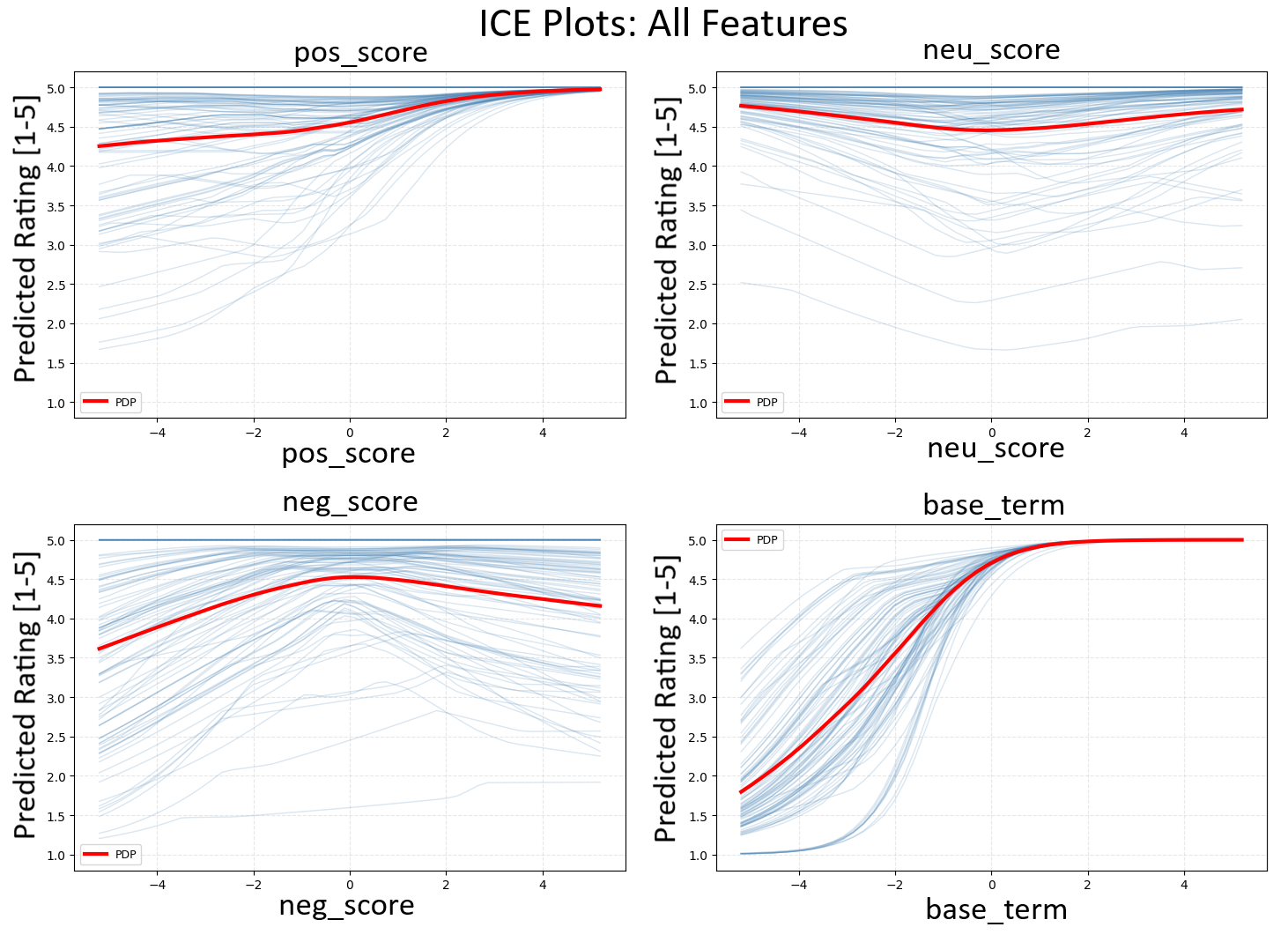}}
	\caption{ICE plots for Amazon dataset (blue: individual curves; red: PDP average).}
	\label{fig:xai_ice}
\end{figure}

The ICE curves (Fig. \ref{fig:xai_ice}) corroborate these findings: \texttt{base\_term} shows a steep sigmoidal PDP with minimal heterogeneity, confirming its dominant role, while $\mathit{score}_{pos}$ trends positively but with some heterogeneity at lower values, suggesting moderation by feature context.

Interestingly, $\mathit{score}_{neg}$ exhibits a non-linear inverted-U shape in its ICE plot, indicating that predictions peak when $\mathit{score}_{neg} \approx 0$. This indicates once more that the model interprets the \textit{complete absence} of negative sentiment as an active positive signal, rather than applying a simple linear penalty. The neutral score ($\mathit{score}_{neu}$) contributes negligibly to the overall prediction, consistent with the SHAP summary plot.

Collectively, these results confirm interpretable behavior: the CF prior anchors predictions, positive sentiment reinforces them, and negative and neutral signals act in a nuanced, non-linear fashion.

\section{Conclusion}
\label{sec:conclusion}
 
In this work, we presented SAGA-CDR, a two-phase emotionally guided CDR framework for the books-to-music setting. The 1\textsuperscript{st} phase combines contextual transformer-based sentiment representations with a stochastic, mask-conditioned CGAN cross-domain mapping and a bias-aware rating prediction module. The 2\textsuperscript{nd} phase uses an LLM to classify books into VA emotional quadrants and filters personalized music candidates to match the book's VA quadrant. Experiments demonstrated that SAGA-CDR excels on prediction accuracy and remains competitive on ranking metrics, while also proving to be highly robust under complex  cross-lingual transfer conditions. Also, SHAP and ICE analyses confirm interpretable, domain-consistent behavior. To the best of our knowledge, this is the first framework to address personal taste and emotional alignment together in a books-to-music CDR setting.

Several promising directions for future work emerge from this study. Firstly, transitioning from discrete emotional quadrants to a continuous VA space could capture fine-grained affective nuances, potentially allowing for chapter-level granularity that produces a dynamic soundtrack evolving alongside the narrative arc. On the audio front, incorporating pre-trained music understanding models to extract embeddings directly from raw waveforms would expand coverage beyond metadata-linked catalogues and provide richer affective estimates. Furthermore, although CGAN training converged stably here, diffusion models, which avoid adversarial optimization altogether \cite{diffusion_rec_model}, could provide a more stable mapping conditioned on source-domain embeddings. For future work, we will also conduct live user studies with a larger population to measure emotional congruence during real reading sessions; the present work establishes only quadrant agreement, since the subjectivity of affective judgments makes perceived congruence measurable only through a controlled study. Finally, we will adapt and evaluate more recent sentiment-aware CDR methods under our formulation.

\section*{Acknowledgment}
This research has been co‐financed by the European Union and Greek national funds through the Program COMPETITIVENESS under the call RESEARCH – INNOVATE  (project code: EK$\Pi$A02-0055774)

\bibliographystyle{IEEEtran}
\bibliography{submission}

\end{document}